\documentclass{article}
\usepackage{spconf,graphicx,cite}
\usepackage[tight,footnotesize]{subfigure}
\usepackage{amsmath,amssymb,amsfonts}
\usepackage{multirow}
\usepackage{multicol}
\usepackage{makecell}

\def\Co{{$R_{\text{\emph{co}}}$}}

\def\C{{$C$}}

\title{Human-Machine Collaborative Video Coding Through Cuboidal Partitioning}
\name{Ashek Ahmmed{$^{1,4}$}, Manoranjan Paul$^{1}$, Manzur Murshed$^{2}$, and David Taubman{$^{3}$}}
\address{{$^{1}$}School of Computing and Mathematics, Charles Sturt University, Australia.\\
$^{2}$School of Science, Engineering, and Information Technology, Federation University, Australia.\\
$^{3}$School of Electrical Engineering and Telecommunications, University of New South Wales, Australia.\\
$^{4}$School of Engineering and Information Technology, University of New South Wales, Australia.}
\begin{document}
%
\maketitle
\begin{abstract}
Video coding algorithms encode and decode an entire video frame while feature coding techniques only preserve and communicate the most critical information needed for a given application. This is because video coding targets human perception, while feature coding aims for machine vision tasks. Recently, attempts are being made to bridge the gap between these two domains. In this work, we propose a video coding framework by leveraging on to the commonality that exists between human vision and machine vision applications using cuboids. This is because cuboids, estimated rectangular regions over a video frame, are computationally efficient, has a compact representation and object centric. Such properties are already shown to add value to traditional video coding systems. Herein cuboidal feature descriptors are extracted from the current frame and then employed for accomplishing a machine vision task in the form of object detection. Experimental results show that a trained classifier yields superior average precision when equipped with cuboidal features oriented representation of the current test frame. Additionally, this representation costs $7\%$ less in bit rate if the captured frames are need be communicated to a receiver.
\end{abstract}
\section{Introduction}
\label{sec:intro}

Many video analysis systems employ a client-server architecture where video signals are captured at the front-end devices and the analyses are carried out at cloud-end servers. In such systems, video data need be communicated from the front end to the server. Prior to this transmission, captured video signal is encoded using a video coding paradigm and at the receiving end the obtained signal is decoded before performing any analysis task. An alternative approach is to extract features from the video data at the client end and then communicate the encoded features rather than the video signal itself. Since capturing of the signal and feature extraction are conducted at the front end, the server becomes less clumsy pertaining to large scale applications and can dedicate most of its resources for the analysis work. However, as the estimated features are devised for specific tasks only, they can be difficult to generalize to a broad spectrum of machine vision tasks in the cloud end.

Traditional video coding standards like H.264/AVC~\cite{AVC} and HEVC~\cite{HEVC} are block-based. They are also pixel and frame centric. In order to model the commonality that exists within a video sequence, the frame that need be coded, known as the current frame, is artificially partitioned into square or rectangular shape blocks. These blocks are formed by grouping together neighboring pixels. After this partitioning, for each current frame block, commonality modeling is employed to form a prediction of it either using the already coded neighboring blocks belonging to the current frame (intra-prediction) or using motion estimation and compensation within a co-located neighborhood region in the set of already coded reference frame(s) (inter-prediction), by minimizing a rate-distortion criterion.

Due to the requirements for communicating feature descriptors, coding standards like compact descriptors for visual search (CDVS)~\cite{CDVS} and compact descriptors for video analysis (CDVA)~\cite{CDVA} were developed. In CDVS, to represent the visual characteristics of images, local and global descriptors are designed. Deep learning features are employed in CDVA to further augment the video analysis performance. Although these features showed excellent performance for machine vision tasks, it is not possible to reconstruct full resolution videos for human vision from such features. This results into two successive stages of analysis and compression for machine and human vision~\cite{VCM}.

As modern video coding standards like HEVC focus on human vision by reconstructing full resolution frames from the coded bitstream and standards like CDVS and CDVA focus on machine vision tasks and incapable of reconstructing full resolution pictures from the coded feature descriptors; there is a need for human-machine collaborative coding that can leverage on the advances made on the frontiers of video coding and feature coding as well as bridge the gap between these two technologies. For example, in autonomous driving use case, some decisions can be made by the machine itself and for some other decisions humans would interact with machine. Such collaborative coding problem is known as video coding for machines (VCM)~\cite{VCM}. In this direction of work, it was proposed in~\cite{VCM-image-coding} to extract features in the form of edge maps from a given image which is used to perform machine vision tasks. After that a generative model was employed to perform image reconstruction based on the edge maps and additional reference pixels.

Although the objectives of human vision and machine vision are different, human-machine collaborative coding can benefit from exploring the significant overlap that exists between these two domains. One important similarity is that human vision is mostly hierarchical. For example, an image of a study room can be described by humans using annotations like - over the floor carpet (global context) there is a desk (local detail) and a laptop (finer detail) is on the desk; on the wall (global context) a clock (local detail) is mounted, etc. Machine vision tasks like object detection can also be modeled in this kind of hierarchical way. For instance, in~\cite{hierarchical} a method is proposed for performing hierarchical object detection in images through deep reinforcement learning~\cite{mnih2015}. Using scale-space theory~\cite{Scale-space} and deep reinforcement learning, another work~\cite{CT} explored optimal search strategies for locating anatomical structures, based on image information at multiple scales.

Murshed \emph{et al.} proposed the hierarchical cuboidal partitioning of image data (CuPID) algorithm in~\cite{cuboid-dicta1}. In CuPID framework, an entire image is initially partitioned into two rectangular shaped regions, known as cuboids, by finding a hyperplane orthogonal to one of the axes. A greedy optimization heuristic, equipped with sum of the information entropy in the split-pair cuboids as the objective function, is employed to find the minimizing hyperplane. This ensures that the obtained split-pair cuboids are maximum dissimilar in terms of image pixel intensity distribution. Next, each cuboid is recursively partitioned by solving the same greedy optimization problem. It was shown that cuboids have excellent global commonality modeling capabilities, are object centric and computationally efficient~\cite{cuboid-dicta2,cuboid-csvt,ashek-icassp,ashek-icassp2021}.

The aforementioned properties of cuboids were leveraged on to traditional video coding in the work of~\cite{ashek-mmsp,ashek-icip2021}; wherein a reference frame is generated using cuboidal approximation of the target frame and incorporating this additional reference frame into a modified HEVC encoder outperformed the rate-distortion performance of a standalone HEVC encoder. The obtained gain in delta PSNR and savings in bit rate can be attributed to the fact that HEVC employs a partitioning scheme that begins at a rigid fixed size of $64\times64$ pixels level and during this process does not take into account the structural properties of the scene need be coded. This sub-optimal partitioning was improved by incorporating cuboids, which are estimated considering the entire frame's pixel intensity distribution and therefore more object-centric.

Building on the work~\cite{ashek-mmsp}, in this paper we investigate the applicability of cuboidal features in video coding targeting machines. Based on the CuPID algorithm, \emph{(i)} features are extracted from a given video frame for the machine vision task of object detection. These features costs less bits to encode. Addition to this, \emph{(ii)} it is possible to reconstruct a full-resolution frame from the obtained feature descriptors alone; unlike the approach in~\cite{VCM-image-coding} that require additional reference pixels along with edge based feature descriptors in this regard. The reconstructed frame is a coarse representation of the original frame and is capable of preserving important structural properties of the original frame. Finally, \emph{(iii)} the object detection performance of the cuboidal descriptors set are put to test across varying number of cuboids.

The rest of the paper is organized as follows: in section II, we briefly describe the cuboidal feature extraction process from a given video frame and the reconstruction process of a full-resolution frame from those estimated feature descriptors. Section III, describes the performance of the cuboidal features over an object detection task. Finally, in section IV, we present our conclusions.

\begin{figure}[t]
\centering
\includegraphics[width=6.5cm,height=5.5cm]{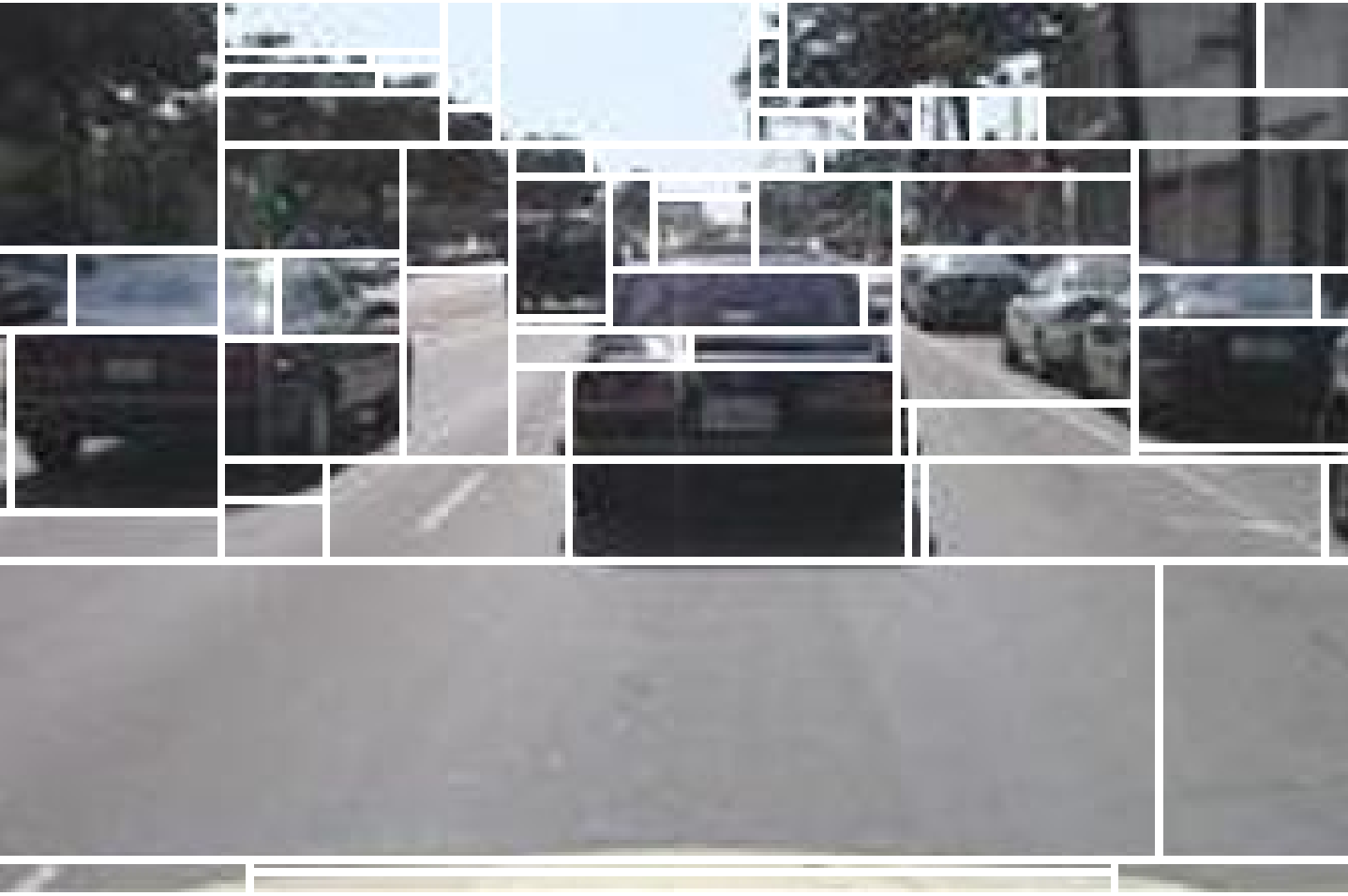}
\caption{The estimated cuboid map over the current frame, \C. This frame is part of the vehicle detection dataset in~\cite{dataset}. The cuboid map, consisting of $70$ cuboids, is determined by the CuPID algorithm \cite{ashek-mmsp}. Each cuboid's coverage is depicted with white boundary pixels.}
\vspace{-4.5mm}
\label{fig:Cn-70}
\end{figure}

\begin{figure*}[t]
\centering
\subfigure[\Co~frame ($n=100$): $R_{\text{\emph{co}}}^{(n=100)}$ ]{\includegraphics[width=0.55\columnwidth,height=4.5cm]{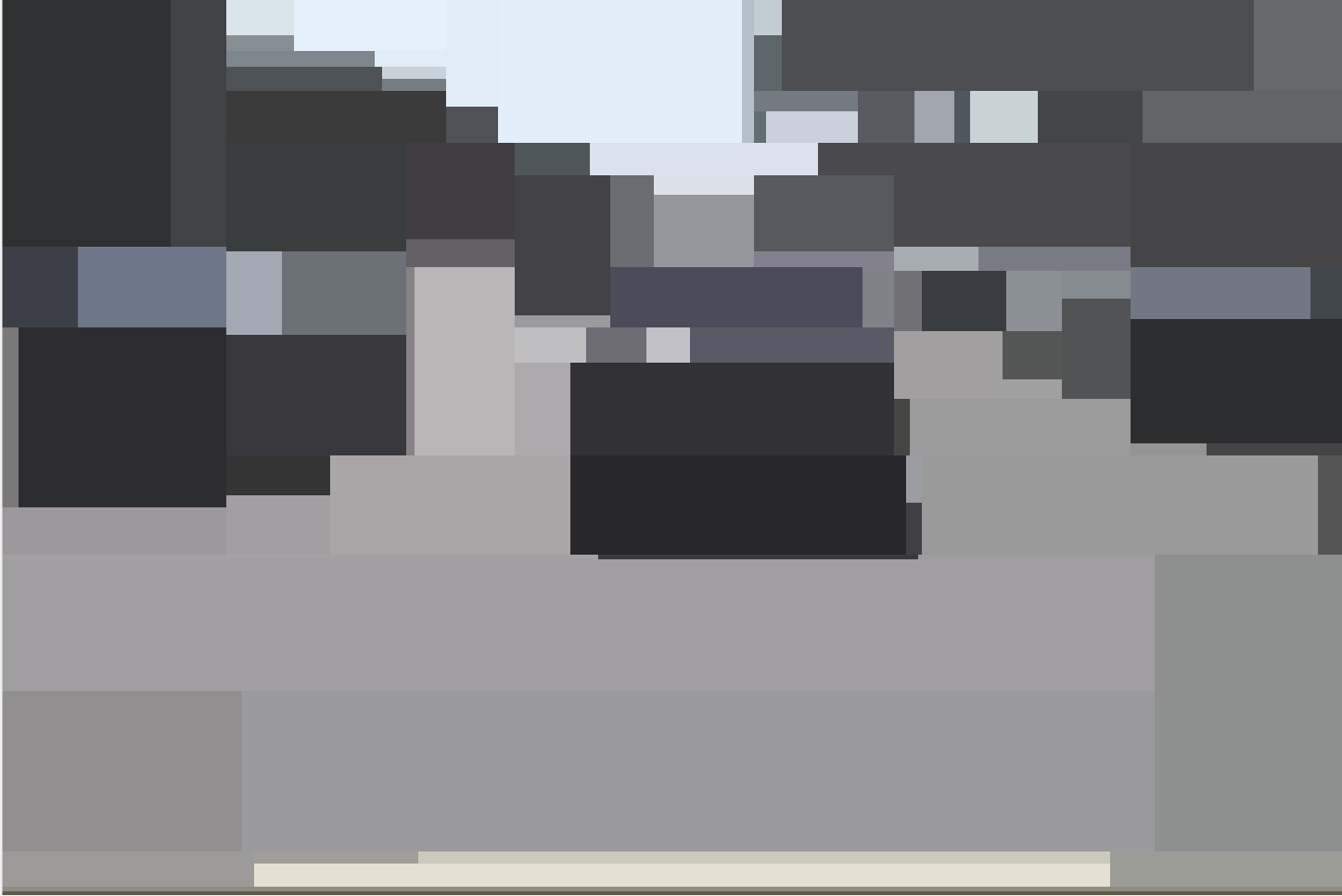}\label{fig:co1}}
\subfigure[\Co~frame ($n=200$): $R_{\text{\emph{co}}}^{(n=200)}$ ]{\includegraphics[width=0.55\columnwidth,height=4.5cm]{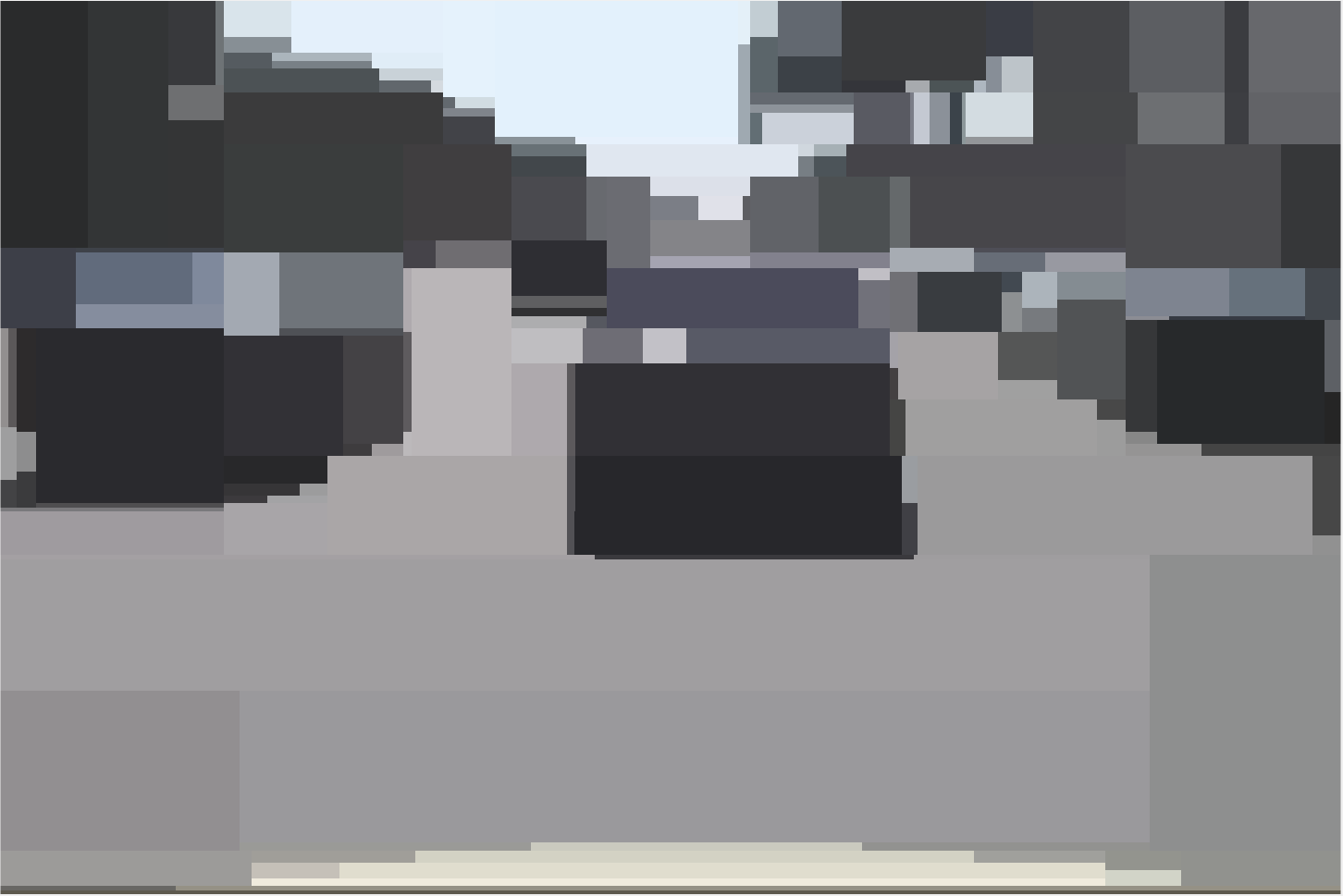}\label{fig:co2}}
\subfigure[\Co~frame ($n=300$): $R_{\text{\emph{co}}}^{(n=300)}$ ]{\includegraphics[width=0.55\columnwidth,height=4.5cm]{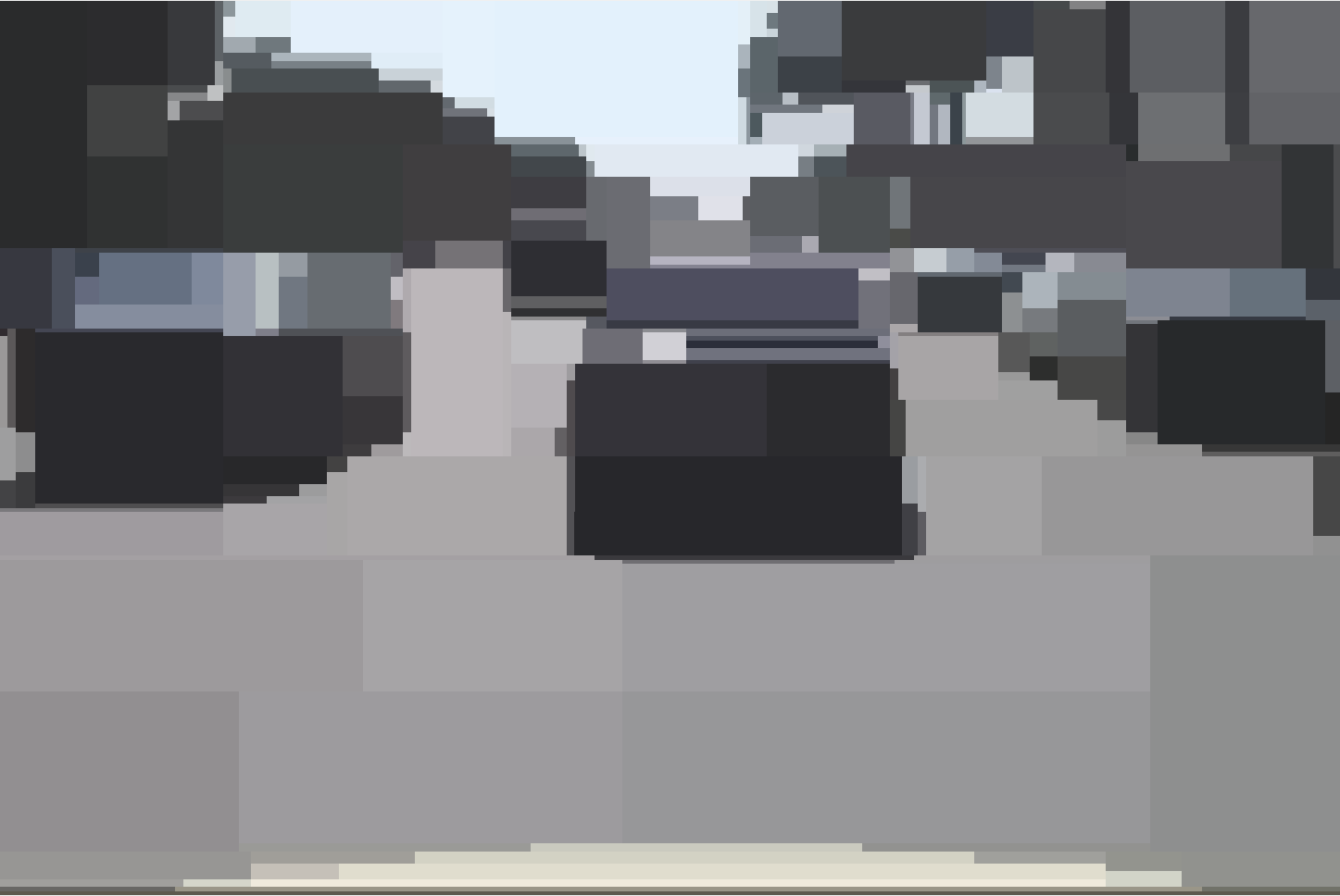}\label{fig:co3}}
\caption{Different coarse representations of the current frame \C. Each one is a \Co~ frame, obtained by varying the number of cuboids, $n$.}
\label{fig:coarse}
\vspace{-4.5mm}
\end{figure*}

\section{Determination of Cuboidal Features from the Current Frame using CuPID}
\label{sec:sum}
Given the original uncompressed current frame \C, of resolution $X\times Y$ pixels, the CuPID algorithm hierarchically partitions it into $n$ cuboids $C^{(1)}, C^{(2)}, \ldots, C^{(n)}$, where $n$ is a user-defined number of tiles. The first partitioning can produce two half-cuboids $C^{(1)}_i$ and $C^{(2)}_i$ of size $i\times Y$ and $(X-i)\times Y$ pixels respectively. This is achieved by selecting a vertical line $x=i+0.5$ over \C~ from the set of lines $i\in\{1,2,\ldots,X-1\}$. Alternatively, \C~ could be split into two half-cuboids $C^{(1)}_{x-1+j}$ and $C^{(2)}_{x-1+j}$ of size $X\times j$ and $X\times (Y-j)$ pixels, respectively by selecting a horizontal line $y=j+0.5$ from the set of lines $j\in\{1,2,\ldots,Y-1\}$.

The first optimal split $s^*$ of \C~ from the possible number of splits $(X-1)+(Y-1)=X+Y-2$ is obtained by solving a greedy optimization \cite{coreman} problem whose objective function is taken to be a measure of  dissimilarity between the candidate cuboid split-pair \cite{ashek-mmsp}. After that each obtained half-cuboid is further split recursively until the total number of cuboids obtained from \C~ becomes equal to the user input $n$. Fig. \ref{fig:Cn-70} shows an example cuboid map for the frame \C. It can be observed that the estimated cuboids are fairly homogeneous in terms of pixel intensity distribution and they occupy arbitrary shaped rectangular regions. Classes like road, sky and trees are represented using separate larger cuboids and important cues for vehicle detection are also preserved.

The estimated cuboid map can be reconstructed at a decoder side from the optimal partitioning indices $\{s^*\}_{i=1}^{n-1}$ alone. That means, for example, if the first optimal partitioning, which split the entire image into $2$ cuboids, took place at height (or width) $h$, the number $h$ need be encoded and so on. These indices are encoded and augmented into a bitstream in the way described in \cite{ashek-mmsp}. Next, for each obtained cuboid, $C^{(i)}$, from the cuboid map, a feature descriptor, $\mathbf{m}^{(i)}$, is computed. This value is taken to be the mean pixel intensity (for each image channel of \C), considering all the pixels $p$ and their intensities $p(x,y)$ within the coverage of the corresponding cuboid.
\begin{equation}
    m_{j}^{(i)} = \frac{\sum_{p\in C^{(i)}}p_{j}(x,y)}{|C^{(i)}|}
\end{equation}
where $j$ represents different color component channels of the frame~\C. The feature descriptors $\mathbf{m}^{(i)}\in{\Bbb Z}^{n}_{*}$ are encoded in the way described in \cite{ashek-mmsp} and are also part of the bitstream.

Having decoded the optimal partitioning indices and the feature descriptors from the communicated bitstream, it is possible to generate a full-resolution frame, \Co. This frame \Co~ provides a coarse representation of the current frame, \C~ since it is created by replacing every encompassing pixel $p\in C^{(i)}$ intensity by the associated cuboidal feature descriptor $\mathbf{m}^{(i)}$. Fig. \ref{fig:coarse} shows examples of this coarse representation frame \Co. It can be observed that this low-frequency representation \Co~ attempt to capture important structure information about the scene and the degree of detailed information it can communicate tends to increase with growing number of cuboids, $n$. Table \ref{tab:time} reports the bits requirements, computational complexity and quality (in terms of PSNR) of these \Co~frames. Here, $b_{c}$ and $t_{c}$ stand for bits and computational time (in seconds) required for an HEVC encoder \cite{HM} to intra-code (with QP $47$) the frame \C. The employed system configuration is: Intel Core i7-8650U CPU@1.90 GHz, 32.0 GB RAM. The maximum PSNR is achieved from the frame $R_{\text{\emph{co}}}^{(n=300)}$ and this frame can be encoded at less than $4\%$ bits requirements that of HEVC. Moreover, the encoding process is $3.45\%$ times faster compared to the HEVC reference software \cite{HM}.

\begin{table}[t]
\begin{center}
\caption{Savings in bits and complexity requirements from the cuboidal feature descriptors at different scale (varying number of cuboids, $n$) over HEVC reference. It also reports the corresponding reconstructed \Co~frames PSNR with respect to the current frame \C.}
\begin{tabular}{ |c|c|c|c|c| }
 \hline
 $\textbf{n}$ & $\textbf{100}$ & $\textbf{200}$ & $\textbf{300}$\\
 \hline
 Savings in bits:$\frac{(b_{c}-b_{co})}{b_{c}}$ & $71\%$ & $38\%$ & $4.05\%$\\
 \hline
 Savings in comp. time:$\frac{(t_{c}-t_{co})}{t_{c}}$ & $72.41\%$ & $41.38\%$ & $3.45\%$\\
 \hline
 Y-PSNR (in dB) & $32.20$ & $32.69$ & $33.16$ \\
 \hline

\end{tabular}
\label{tab:time}
\vspace{-0.5cm}
\end{center}
\end{table}

\section{Object Detection using the Cuboidal Feature Descriptors}
\label{sec:sum}
The cuboidal feature descriptors attempt to capture important structural properties of the image at comparatively lower bit and computational complexity requirements. In this work, we propose to utilize them for a machine vision task, in particular the \Co~frames are employed in a vehicle detection problem. For the addressed vehicle detection problem, the object detector You only look once (YOLO) v2~\cite{yolo} is used. YOLO v2 is a deep learning object detection framework that uses a convolutional neural network (CNN)~\cite{cnn} for detection task. Using unoccluded RGB images of the front, rear, left, and right sides of cars on a highway scene, the detector is trained. The CNN used with the vehicle detector, uses a modified version of the MobileNet-v2 network architecture~\cite{mobnet}. $60\%$ of the data is used for training, $10\%$ for validation, and the rest $30\%$ is used for testing the trained detector.

\begin{figure}[t]
\centering
\subfigure[HEVC coded test frame with detection confidence scores: $\{0.79, 0.80, 0.73, 0.81\}$ respectively.]{\includegraphics[width=0.49\columnwidth,height=4.5cm]{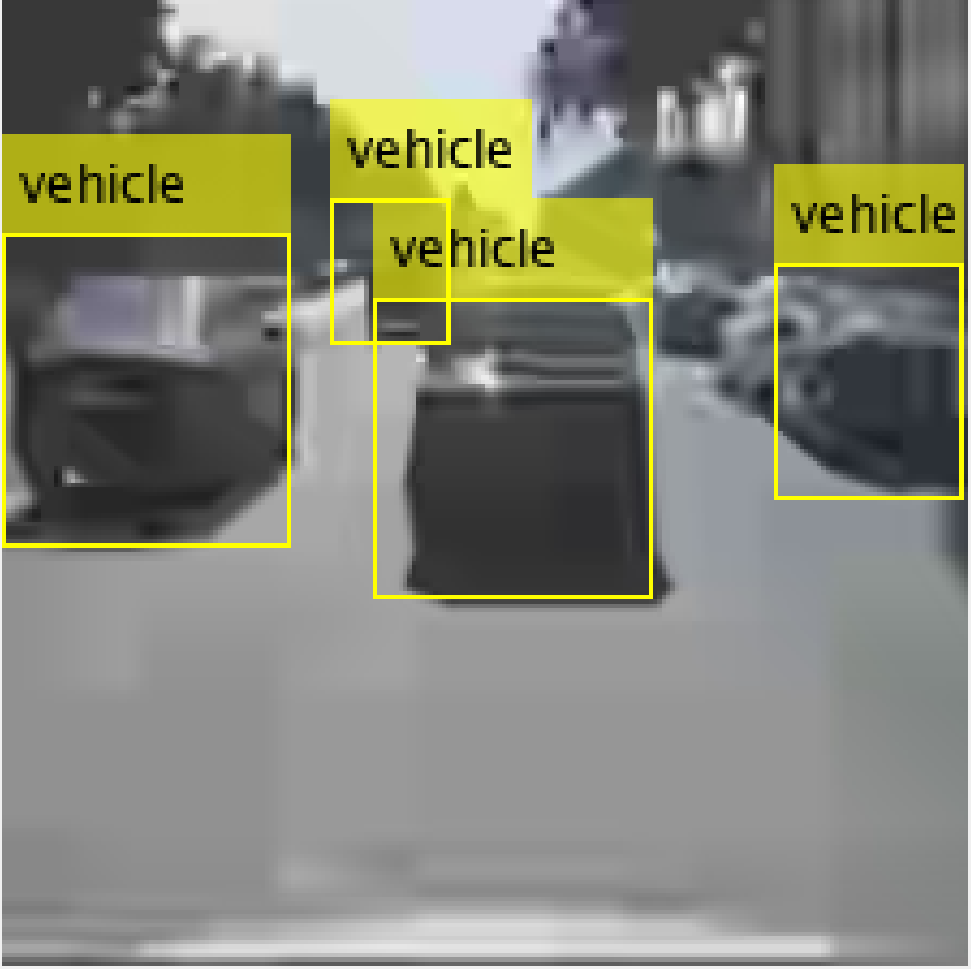}\label{fig:hevc-d}}
\subfigure[$R_{\text{\emph{co}}}^{(n=300)}$ test frame with detection confidence scores: $\{0.78, 0.68, 0.78, 0.86\}$ respectively. ]{\includegraphics[width=0.49\columnwidth,height=4.5cm]{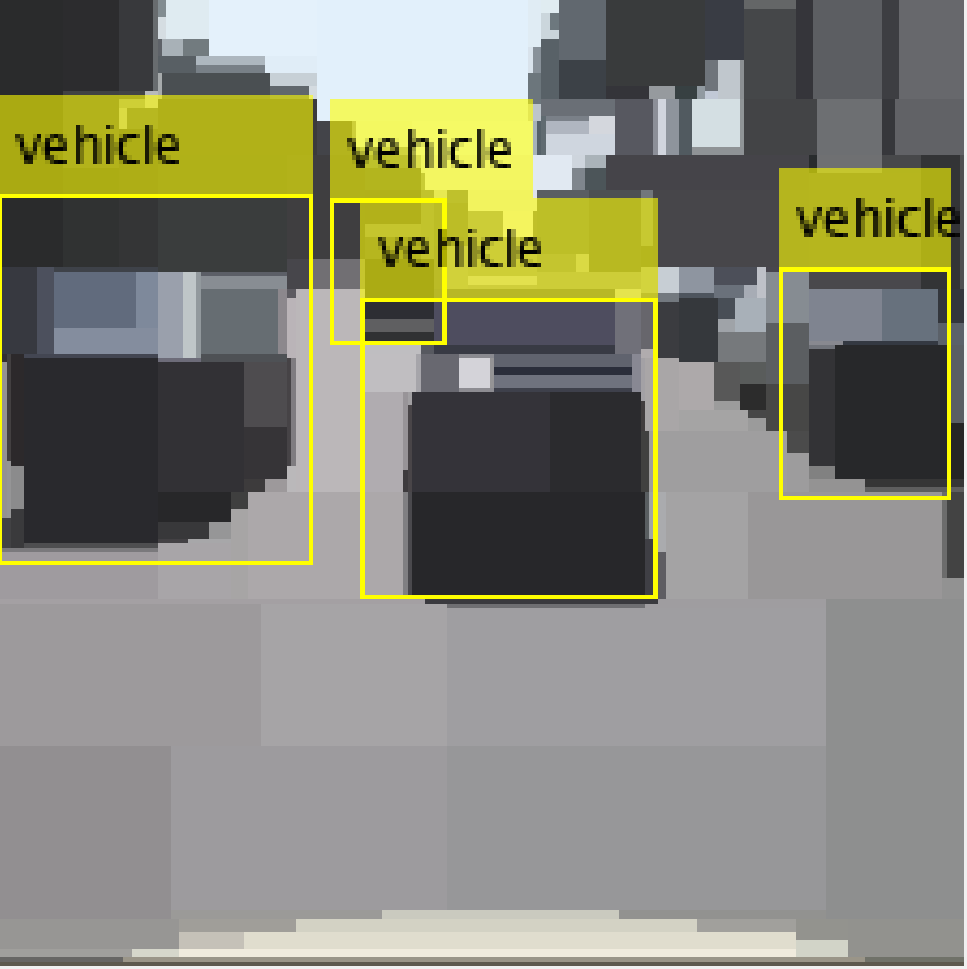}\label{fig:Rco-300-d}}
\caption{The trained vehicle detector's performance over different test frame. (From left to right) HEVC intra-coded version and coarse representation \Co~ of \C~respectively. The detector managed to detect same number of vehicles in the \Co~frame with various confidence scores.}
\label{fig:coarse}
\vspace{-4.5mm}
\end{figure}

\begin{table}[t]
\centering
\caption{Average precision (AP) values of the trained vehicle detector over different representations of the same test set \cite{dataset}. It also reports the bit rate required to code the test set frames.}
\begin{tabular}{ |c|c|c| }
 \hline
 Testset & AP & Bit rate (Kbps)\\
 \hline
 HEVC coded & $0.9366$ & $110.63$\\
 \hline
 $R_{\text{\emph{co}}}^{(n=100)}$ & $0.8158$ & $31.91$\\
 \hline
 $R_{\text{\emph{co}}}^{(n=200)}$ & $0.9083$ & $67.53$\\
 \hline
 $R_{\text{\emph{co}}}^{(n=300)}$ & $\textbf{0.9462}$ & $102.74$\\
 \hline
\end{tabular}
\label{tab:results}
\vspace{-0.5cm}
\end{table}

Once the detector has been trained, it is used to detect vehicles over \emph{(i)} HEVC coded versions of the original test video frames (refers to Fig. \ref{fig:hevc-d}) and \emph{(ii)} cuboidal descriptor oriented coarse representation frames (\Co) of the original test frames (refers to Fig. \ref{fig:Rco-300-d}). As can be seen, the detector managed to detect same number of vehicles in the $R_{\text{\emph{co}}}^{(n=300)}$ frame like its HEVC coded counterpart. Moreover, for $2$ vehicles the detector's detection performance (in terms of higher confidence score) improved when compared to the HEVC coded frame. However, for the second vehicle, the confidence score is lower in the case of $R_{\text{\emph{co}}}^{(n=300)}$ frame. This is because that vehicle in the scene is partly occluded by another vehicle and therefore its cuboidal representation lacks in some structural information.

\begin{figure}[t]
\centering
\includegraphics[trim=0cm 0.5cm 0 1.5cm, clip,width=0.95\columnwidth]{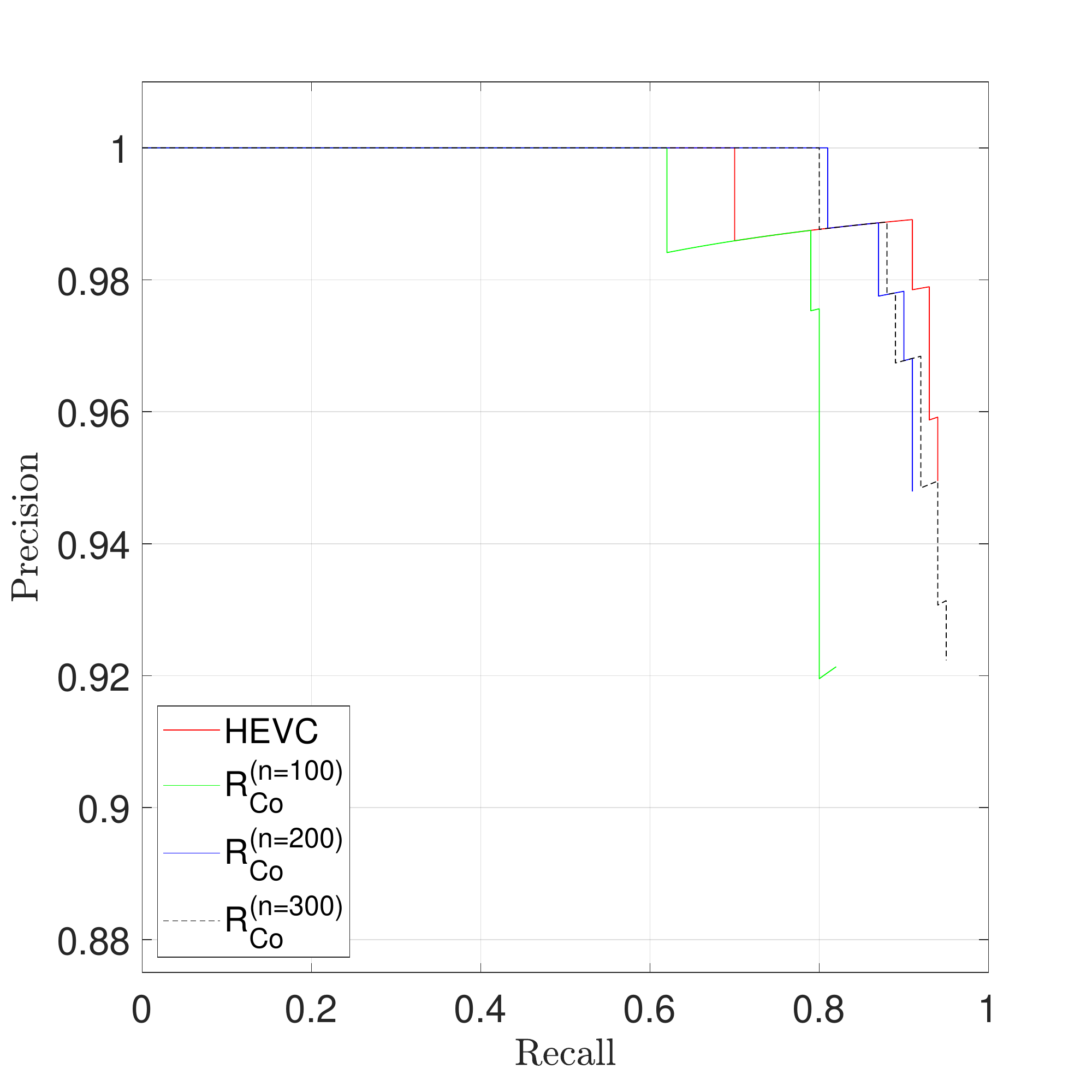}
\caption{Precision recall performance of the trained vehicle detector over different representations of the same test set.}
\vspace{-4.5mm}
\label{fig:PRCurve}
\end{figure}

Table \ref{tab:results} reports the performance of the detector over these different test data in the form of average precision. It can be noted that the detector produced the maximum average precision of $0.9462$ over $R_{\text{\emph{co}}}^{(n=300)}$ test frames. This result is achieved with a bit rate requirements that of HEVC; in particular a bit rate savings of around $7\%$ is obtained. Bit rate savings improve further (a savings of $39\%$) if the $R_{\text{\emph{co}}}^{(n=200)}$ test frames are used instead. However, in this case the average precision decreases ($-0.0283$) compared to the HEVC coded test set. Fig. \ref{fig:PRCurve} shows the precision-recall curves for the employed test sets. At almost $80\%$ recall, the precision is more over the test sets $R_{\text{\emph{co}}}^{(n=200)}$ and $R_{\text{\emph{co}}}^{(n=300)}$, compared to the HEVC coded test set. And these cuboidal descriptor oriented test sets take less bits to encode than HEVC.

\section{Conclusion}
Despite the difference in objectives, there exists similarities between human vision and machine vision. A collaborative coding paradigm is required for many use cases. The cuboidal descriptors oriented representation of the current frames were shown to be beneficial in traditional video coding. Leveraging on their properties like object oriented, scene structural awareness, compact representation and computational simplicity, in this paper, cuboidal descriptors are employed in a machine vision task. Experimental results show that the addressed vehicle detection problem can be solved more accurately (average precision improved by $0.01$) if cuboidal descriptors are used. Along with this a savings in bit rate of around $7\%$ is reported over a HEVC reference.





\bibliographystyle{IEEEbib}
\bibliography{refs}

\end{document}